\documentclass[a4paper, twocolumn]{article}

\usepackage{amsmath}
\usepackage{graphicx}
\usepackage{overcite}
\usepackage{abstract}
\usepackage{url}

\renewcommand\citeform[1]{[#1]}

\title{\Large\bf Effective system for simulating dust continuum observations \\
on distributed computing resources }

\author{Kazutaka MOTOYAMA\thanks{National Institute of Informatics, 2-1-2 Hitotsubashi, Chiyoda-ku, Tokyo 101-8430, Japan}, 
    Yoshikazu TANAKA\thanks{Meisei University, 2-1-1 Hodokubo Hino, Ome, Tokyo 191-8506, Japan}, Kento AIDA$^*$, \\ 
Eisaku SAKANE$^*$, and Kenichi MIURA$^*$}

\date{}

\begin{document}
\twocolumn[
    \begin{@twocolumnfalse}
    \maketitle
    \begin{abstract}
        {\normalsize We present an effective system for simulating dust continuum observations by radiative transfer 
            simulations.  
            By using workflow management system RENKEI-WFT, we utilized distributed 
            computing resources and automated a sequence of computational tasks required for radiative 
            transfer modeling, namely, main radiative transfer simulations, pre-/post-processes, and data transfer 
            between computing resources.  
            Our system simultaneously executes a lot of radiative transfer simulations with different input parameters 
            on distributed computing resources. 
            This capability of our system enables us to conduct effective research by radiative transfer simulation.
            As a demonstration of our system, we simulated dust continuum observations of protoplanetary disk.
            We performed hydrodynamic simulation modeling photoevaporating protoplanetary disk irradiated by 
            ultra violet radiation from nearby massive stars. 
            Results of this hydrodynamic simulation were used as input data for radiative transfer simulations.
            Expected spectral energy distributions and intensity maps were obtained by our system.\\
            \\
            \textbf{Keywords: }distributed computing, radiative transfer, scientific workflow
            \\
        }
    \end{abstract}
  \end{@twocolumnfalse}
]
\saythanks

\section{Introduction}
Predicting radiative properties of astronomical objects is essential to interpretation of photometric and 
spectroscopic observations.
Spectral line radiation from molecular gas and continuum radiation from dust grain  
provide useful information on physical and chemical properties of observed objects.
Recent increasing of computational power allows us to simulate observations of astronomical objects by 
radiative transfer simulation. 
Comparison between observations and radiative transfer model helps understanding of physical and 
chemical properties of observed objects. 
Thanks to its high sensitivity and resolution, 
future powerful facilities such as the Atacama Millimeter Array (ALMA) and 
the Thirty Meter Telescope (TMT) will provide huge observational data at radio/infrared/optical wavelength.
It is necessary to establish effective method for radiative transfer modeling to make comparisons 
with these huge observational data. 

Automation of computational tasks for radiative transfer modeling, such as main 
radiative transfer simulations and pre-/post-processes,  
is needed to improve efficiency of radiative transfer modeling.
Since radiative transfer modeling for simulating observations involves observing direction and 
observing frequency as parameter, 
we need to run a lot of radiative transfer simulations with different parameters. 
It makes manual operation of computational tasks inefficient.

When all computational tasks for simulating observations run on single computing resource, 
shell script provides us an easy way to automate these computational tasks.  
However, it is very common situation that main simulations and pre-/post-processes run on different computing resources. 
In such case, we encounter difficulty to automate computational tasks with shell script. 
We propose using a workflow management system RENKEI-WFT to automate computational tasks. 
RENKEI-WFT was developed for REsources liNKage for E-scIence (RENKEI) 
project\footnote{\url{http://www.e-sciren.org/}}launched by 
National Institute of Informatics (NII) and partner institutes and universities. 
RENKEI-WFT coordinates multiple job submission and data transfer on distributed computing resources integrated 
with grid middleware such as NAREGI\footnote{\url{http://www.naregi.org/}}. 
One of the advantages of using RENKEI-WFT is that it supports parallel execution of a lot of independent jobs 
with different input parameters. 
Therefore, RENKEI-WFT is suitable for executing parameter survey application such as radiative transfer simulation. 

Other advantage of using RENKEI-WFT is that it ensures secure communications 
between computing resources.
Security is one of key issues in distributed computing environment, particularly when utilizing computing 
resources distributed over multiple organizations. 
RENKEI-WFT uses the Grid Security Infrastructure (GSI) as an authentication mechanism.
GSI is designed based on the X.509 public key infrastructure, where a certificate 
authority digitally signs a user's public key and issues a digital certificate.
The certificate is used for authentication between computing resources.

This paper aims to propose method for automation of simulating observations on distributed 
computing resources with RENKEI-WFT. 
The layout of this paper is as follows. Section \ref{sec:rt} describes our radiative 
transfer simulation. Section \ref{sec:renkei-wft} presents architecture of RENKEI-WFT. 
Section \ref{sec:design} describes design and implementation of our system and demonstrates how it works. 
Finally, section \ref{sec:summary} 
summarizes our work presented in this paper.


\section{Radiative Transfer Simulation}\label{sec:rt}
This section briefly describes radiative transfer simulation that is main component of our system. 
Now we define the specific intensity, $I_{\nu}$, as the amount of radiative energy at frequency $\nu$ 
crossing a surface 
of unit area per unit time, per unit solid angle, per unit frequency interval. 
As described in Rybicki and Lightman (1979)~\cite{1979rpa..book.....R}, the change in specific intensity  
traveling a distance $s$ in a specific direction 
is given by radiative transfer equation:
\begin{equation}
    \frac{d I_{\nu}}{d s} = - \alpha_{\nu}  I_{\nu} + j_{\nu}, 
    \label{eq:transfer}
\end{equation}
where $\alpha_{\nu}$ and $j_{\nu}$ are the absorption 
coefficient and the spontaneous emission coefficient, respectively.
Although both of molecules and dust can be sources of emission and absorption, 
we focus only on transfer of thermal continuum radiation from dust in radiative equilibrium with local 
radiation field in this paper.
Since the equilibrium temperature of dust is set by balance between emitted and absorbed radiation, 
it is written as 
\begin{equation}
    j_{\nu} = \alpha_{\nu} B_{\nu}(T_{dust}),
    \label{eq:equilibrium}
\end{equation}
where $B_{\nu}$ is the Planck function at the dust temperature $T_{dust}$. 
The absorption coefficient is related to the dust opacity per unit mass, $\kappa_{\nu}$, by 
\begin{equation}
    \alpha_{\nu} = \kappa_{\nu} \rho_{dust},
    \label{eq:kappa}
\end{equation}
where $\rho_{dust}$ is the mass density of dust.
The dust opacity depends on the shape, size distribution, and constitution of the dust grain.

Radiative transfer code RADMC-3D\footnote{\url{http://www.ita.uni-heidelberg.de/~dullemond/software/radmc-3d/}} 
is used for radiative transfer simulations in our system.
Simulation of thermal dust continuum radiation consists of two steps. 
At first step, equilibrium dust temperature $T_{dust}$ satisfying equation (\ref{eq:equilibrium}) is 
computed by Monte Carlo method (Lucy 1999~\cite{1999A&A...344..282L}; Bjorkman \& Wood 2001~\cite{2001ApJ...554..615B}). 
The positions, spectra and luminosities of individual stars inside the observed system, and spatial distribution 
of dust density are given as input data.
At second step, expected images and spectra are computed by ray tracing. 
The radiative transfer equation given by equation (\ref{eq:transfer}) is solved along direction pointing to the observer. 

\section{RENKEI-WFT}\label{sec:renkei-wft}

We assume the typical application scenario in which the user utilizes computing resources in both 
a laboratory/department grid (or a local site) and a national grid infrastructure to run the application. 
Hereafter in this paper, we refer to the former resources as a Laboratory Level System (LLS) and the latter 
resources as a National Infrastructure System (NIS). The workflow application consists of jobs for 
pre-/post-processes and the main simulation. The user would want to run jobs for the pre-/post-processes 
on the LLS and the job for the main simulation on the NIS. 

\begin{figure}[tb]
  \centering
  \includegraphics[width=80mm]{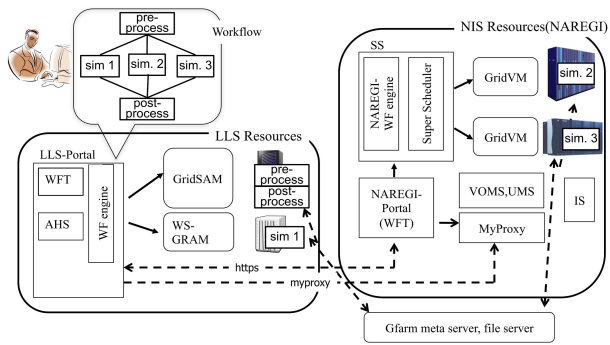}
  \caption{Example of an LLS and a NIS setting}
  \label{fig:slide01}
\end{figure}

RENKEI-WFT enables submission of jobs in a single workflow to both the LLS and the NIS even if 
they are operated with different grid middleware. 
Figure \ref{fig:slide01} illustrates an example of an  LLS and a NIS setting using RENKEI-WFT. 
The LLS consists of an LLS-Portal server and computing resources, e.g., PC cluster(s). 
Here, we assume that two PC clusters in the LLS are operated with different grid middleware, 
GridSAM\footnote{\url{http://www.omii.ac.uk/wiki/GridSAM/}} and WS GRAM\footnote{\url{http://www.globus.org/toolkit/docs/4.0/execution/wsgram/}}. 
Both types of middleware provide lightweight  services for job submission and monitoring of submitted jobs. 
The NIS consists of supercomputers operated through NAREGI. 
NAREGI includes the information service (IS) and the brokering service (Super Scheduler), 
which performs matchmaking and dispatches jobs to suitable computing resources in the NIS.

RENKEI-WFT, which consists of the workflow tool (WFT) and the workflow engine (WF engine), is installed on a portal system in the LLS (LLS-Portal). The WFT works as a user interface for editing a workflow, submitting jobs in the workflow, and monitoring the submitted jobs. In addition, the WFT offers its service on the portal so that the user does not have to install and configure software for RENKEI-WFT in the local machine. The WF engine dispatches jobs through its job submission interfaces. 
The user can save edited workflows in the application repository, Application Hosting Service (AHS) \cite{kanazawa2010}, 
and share the workflows with other users.

RENKEI-WFT is designed to cooperate with the grid file system, Gfarm \cite{tatebe2004}, to enable file sharing among distributed computing resources. Jobs running in the LLS and the NIS are able to share files through Gfarm, which enables the stage-in/stage-out of files between the LLS and the NIS. The interface to access Gfarm is implemented by the GridFTP protocol \cite{allcock2003} or a proprietary protocol.

Resources in the NIS consist of supercomputers in production service. They are usually operated with rigid firewall policies. Administrators of the NIS may not be able to open special TCP ports for submitting jobs from the LLS. We designed RENKEI-WFT to use limited protocols from the LLS to the NIS. Access from the LLS to the NIS is conducted by using the https request between RENKEI-WFT and NAREGI-Portal.

The detailed scenario presented in Figure \ref{fig:slide01} is as follows. The user accesses the LLS-Portal and launches RENKEI-WFT. Then, the user edits the workflow, which consists of five jobs, \texttt{pre-process}, \texttt{sim1}, \texttt{sim2}, \texttt{sim3} and \texttt{post-process}. The user is able to submit the workflow, or jobs, and through the WFT, the user can monitor the submitted jobs that are running in the LLS-Portal.  The WFT shows statuses, \lq\lq queued", \lq\lq running", or \lq\lq done" of the submitted jobs on the GUI. The WF engine in the LLS-Portal submits jobs to the LLS directly and to the NIS via the NAREGI-Portal using the https protocol. In this example, \texttt{pre-process}, \texttt{post-process}, and \texttt{sim1} run on the LLS, and \texttt{sim2} and \texttt{sim3} run on the NIS. Through the WFT, the user is able to indicate resources (LLS or NIS) to which the jobs are assigned. The rest of this section presents the details of RENKEI-WFT.

\subsection{Workflow Tool}
\label{sec:wft}

\begin{figure}[tb]
  \centering
  \includegraphics[width=80mm]{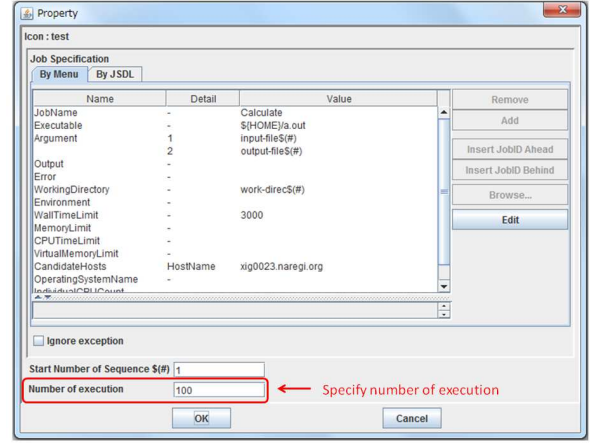}
  \caption{GUI to define the property of a calculate icon}
  \label{fig:snap04}
\end{figure}

The Workflow Tool (WFT) is a core software module in RENKEI-WFT. It enables the user to edit the workflow, 
submit the workflow, and query the status of the workflow (or jobs) running on the LLS and the NIS through the GUI. 
A workflow consists of programs to be executed once, programs to be executed a specified number of times, 
data, and sub-workflows. 
These components are presented as program icon, calculate icon, data icon, and workflow icon 
in the GUI, respectively. 
The user defines the property for each icon. For example, a data icon has a property indicating a file path. 
The property of a program icon includes the path of a program executable, environment variables, and the computing resource to run the program. 
The property of a calculate icon includes a number to specify how many times a program should be executed. 
Working directories and input/output files can be named with a sequential number. 
It is useful for parameter survey application. 
Figure \ref{fig:snap04} shows a screenshot of the GUI for defining the property of a calculate icon. 
A sub-workflow contains other workflows, such that the WFT is able to create a hierarchical workflow structure. Complex workflow descriptions such as loops and conditional branches are also supported. 

\begin{figure}[tb]
  \centering
  \includegraphics[width=80mm]{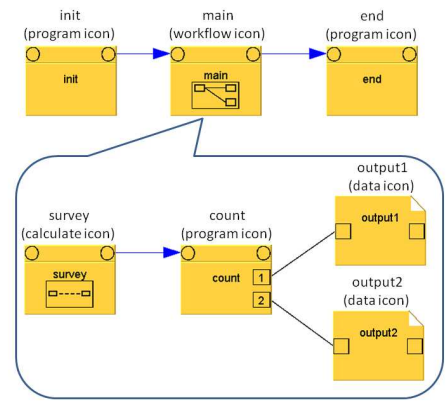}
  \caption{An example of a workflow}
  \label{fig:icons}
\end{figure}

Figure \ref{fig:icons} illustrates an example of a workflow with a two-level structure. The workflow at 
the outermost level is composed of two program icons, \texttt{init} and \texttt{end}, and a workflow icon, \texttt{main}. 
The workflow, \texttt{main}, consists of a calculate icon, \texttt{survey}, and a program icon, \texttt{count}, and 
two data icons, \texttt{output1} and \texttt{output2}. 
The lines and the lines with an arrow denote the data flow and the control flow between two icons, respectively. 
In this example, \texttt{init} runs first, then \texttt{main} runs after \texttt{init} finishes. 
In the execution of \texttt{main}, \texttt{survey} runs first, then \texttt{count} runs after \texttt{survey} finishes and 
writes data into \texttt{output1} and \texttt{output2}. 
Finally, \texttt{end} runs after \texttt{main} finishes.

\subsection{Internal Representation of the Workflow}
\label{sec:internal}

RENKEI-WFT enables users to submit jobs to computing resources operated by different middleware. Thus, an internal representation of the workflow in RENKEI-WFT should be architecture independent among the various middleware used by the computing resources. We designed the intermediate representation of the workflow, the NAREGI-Workflow Markup Language (NAREGI-WFML). NAREGI-WFML is extended from the Grid Service Flow Language (GFSL)~\cite{Krishnan:2002} by adding a sub-workflow representation. The grammar of NAREGI-WFML is architecture independent, and NAREGI-WFML presents a job (or a program) in the Job Submission Description Language (JSDL)~\cite{anjomshoaa2008}. JSDL is standardized in the Open Grid Forum (OGF)\footnote{\url{http://www.ogf.org/}}, and many grid middleware services support JSDL to improve interoperability.

\begin{figure}[tb]
  \centering
  \includegraphics[width=80mm]{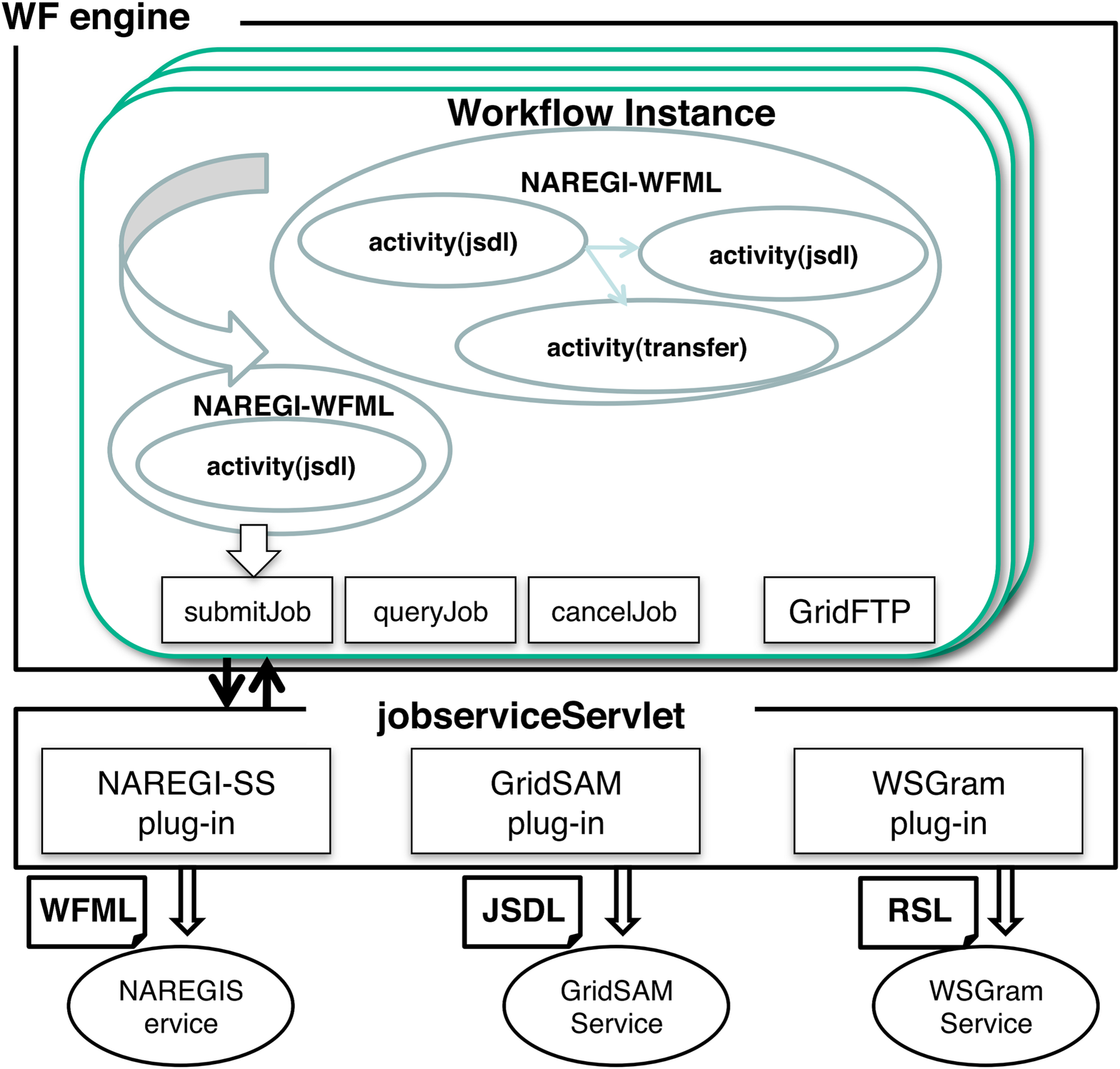}
  \caption{Workflow Engine}
  \label{fig:slide05}
\end{figure}

Figure \ref{fig:slide05} illustrates the structure of the Workflow Engine (WF engine) and the servlet to submit jobs (jobserviceServlet). The WFT creates a workflow instance per each workflow and puts the workflow into the threads pool, that is, multiple workflows are executed in parallel using multithreading. We use the ExecutorService interface in Java for thread management. For each workflow, the WF Engine decomposes the workflow represented by NAREGI-WFML into sub-workflows consisting of computing jobs and file transfer jobs. Then, it submits jobs in the sub-workflows to computing resources in the LLS or the NIS through the jobserviceServlet. The WF Engine creates a sub-workflow of one activity (or one job) for a job submitted to GridSAM or WS GRAM. The GridSAM plug-in module extracts a JSDL document from the NAREGI-WFML document  and submits the job to the GridSAM Service in the LLS. The WS GRAM plug-in module extracts a JSDL document, translates it to a Resource Specification Language (RSL)\footnote{\url{http://www.globus.org/toolkit/docs/2.4/gram/rsl_spec1.html}} document and submits it to the WS GRAM Service in the LLS. NAREGI supports NAREGI-WFML. The WF Engine creates sub-workflows of multiple activities, and the NAREGI plug-in module submits the sub-workflow in NAREGI-WFML to the NAREGI service. File transfer jobs in the LLS are executed directly using the GridFTP protocol. The WF engine monitors the status of submitted jobs by using the polling mechanism. 

\begin{figure}[tb]
  \centering
  \includegraphics[width=80mm]{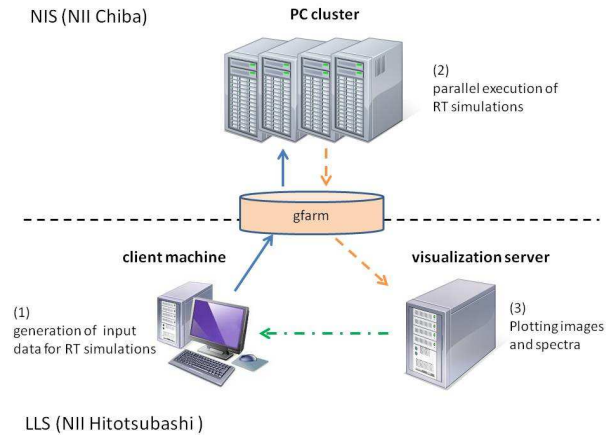}
  \caption{Schematic diagram of system configuration. Arrows indicate directions of data transfer.}
  \label{fig:schematic}
\end{figure}

\subsection{Authentication}\label{sec:auth}
\begin{figure*}[tb]
  \centering
  \includegraphics[width=170mm]{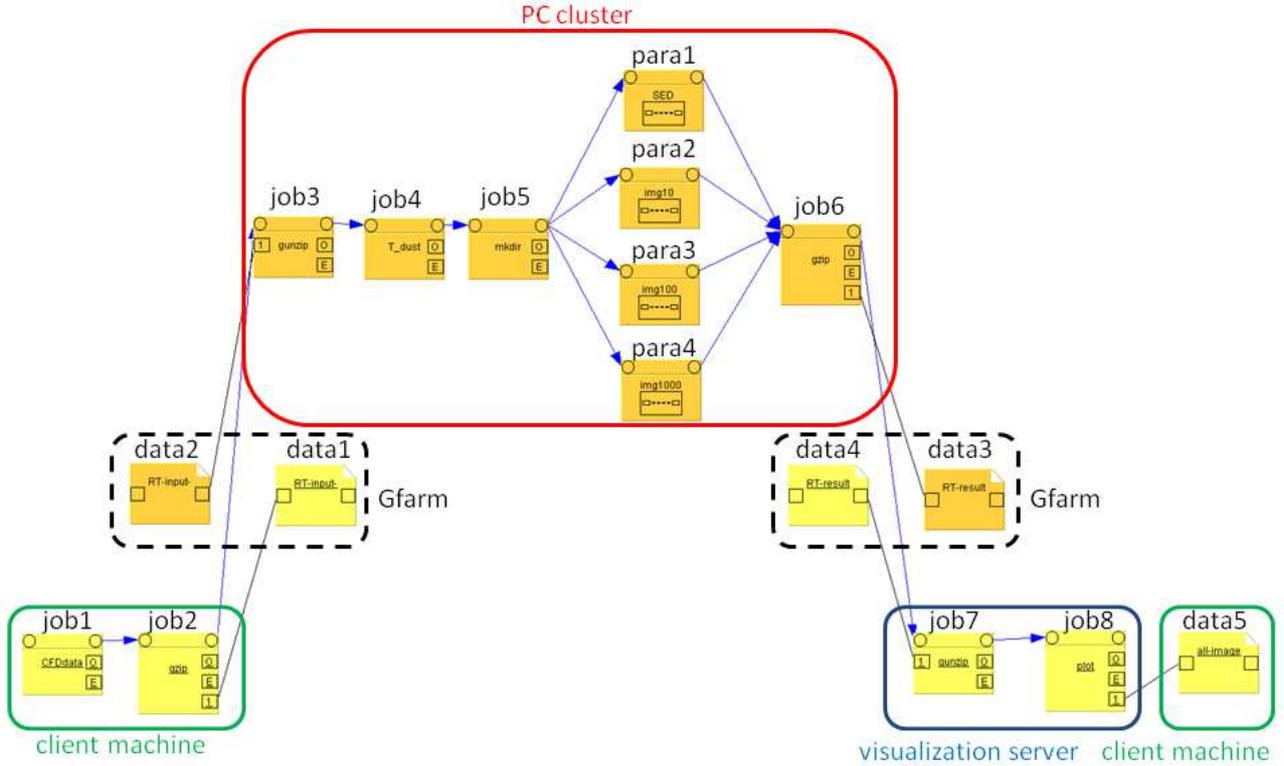}
  \caption{Workflow of our system. Program icons, calculate icons, and data icons are labeled with 
  "job", "para", and "data", respectively.}
  \label{fig:our-wf}
\end{figure*}

Authentication is an important issue when utilizing computing resources distributed over multiple organizations. RENKEI-WFT uses the Grid Security Infrastructure (GSI) as an authentication mechanism for submitting jobs to computing resources in the LLS and the NIS. We use the User Management Server (UMS) in NAREGI to share user certificates between the LLS and the NIS, where user certificates are archived in a certificate repository of the UMS.

When starting to use RENKEI-WFT, the user first creates a proxy credential from the user's certificate archived in the UMS. The proxy credential is stored in the MyProxy server\footnote{\url{http://grid.ncsa.illinois.edu/myproxy/}}. As long as the proxy credential is valid, the user is able to sign on to the LLS-Portal. The LLS-Portal and the NAREGI-Portal are able to obtain the proxy credential from the MyProxy server. When the user submits a job, the proxy credential is delegated to the computing resource that runs the user's job.

The credential includes not only the basic authentication information but also the virtual organization (VO) information issued from the Virtual Organization Membership Service (VOMS)~\cite{alfieri2004}. The VO information is mainly used in the NIS operated by NAREGI.

\section{Design and Implementation}\label{sec:design}
Figure \ref{fig:schematic} shows a schematic diagram of system configuration.
The system is composed of distributed computing resources that are located at NII Chiba site and NII Hitotsubashi site. 
Following application scenario described in section \ref{sec:renkei-wft}, we refer to computing resources 
in NII Chiba site 
and those in NII Hitotsubashi site as NIS and LLS, respectively.
Input data for radiative transfer simulations are generated on client machine in LLS. 
Radiative transfer simulations run on PC cluster in NIS.  
Expected spectra and images are plotted with visualization software 
VisIt\footnote{\url{https://wci.llnl.gov/codes/visit/}} on visualization server in LLS, and generated image files are 
sent to client machine. 
Gfarm file system is used to share files between NIS and LLS.

Figure \ref{fig:our-wf} shows workflow of our system. The workflow consists of data icons, program icons, 
and calculate icons. These icons are labeled with "data", "job", and "para", respectively. 
Workflow in Figure \ref{fig:our-wf} is executed in the following steps: 
\begin{enumerate}

\item Input data for radiative transfer simulations are generated on client machine in LLS (job1).

\item These input data are transferred to PC cluster in NIS via Gfarm. 
    Client machine in LLS compresses data files by tar command with gzip option before transfer (job2), 
    and write these files on gfarm (data1).
    Then, PC cluster in NIS reads these files on gfarm (data2) and decompressed them (job3).

\item Equilibrium dust temperature is computed by Monte Carlo method. 
    RADMC-3D is executed with option "mctherm" (job4).

\item Ray tracing calculations are performed on PC cluster.
    Working directories for each job are constructed (job5). 
    Spectral energy distributions (para1), intensity maps at 10 $\mathrm{\mu}$m (para2), 100 $\mathrm{\mu}$m (para3), 
    and 1000 $\mathrm{\mu}$m (para4) with different viewing angles are computed in parallel.
    RADMC-3D is executed with option "sed" and "image".

\item Results of radiative transfer simulations are transferred to visualization server in LLS in the same manner 
    as step 2 (job6, data3, data4, and job7).

\item Using results of radiative transfer simulations, spectra and images are plotted on visualization server (job8). 

\item All image files generated on visualization server are directly transferred to client machine 
    using the GridFTP protocol (data5). 

\end{enumerate}

\begin{figure}[tb]
  \centering
  \includegraphics[width=80mm]{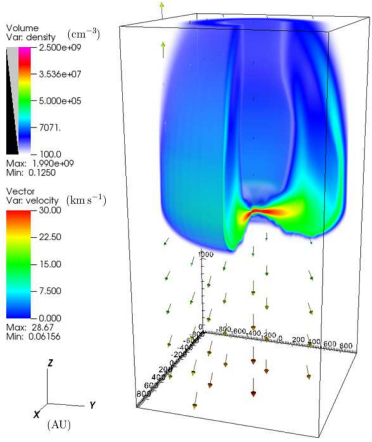}
  \caption{Volume rendering map of neutral gas and velocity field (arrows) at $t = 4.5$ kyr.}
  \label{fig:hydro}
\end{figure}

For demonstration of our system, we simulated observations of photoevaporating protoplanetary disk 
using results of hydrodynamic simulation. 
A number of young stellar objects surrounded by ionized envelope have been detected in H\,{\scshape II} regions. 
These cometary-shaped objects are called "proplyd" and supposed to be photoevaporating 
protoplanetary disks irradiated by strong ultra violet radiation from nearby massive stars. 
We performed hydrodynamic simulation modeling photoevaporating protoplanetary disk. 
We assumed that disk is 1 pc away from exciting star whose mass and surface temperature are 18 $\mathrm{M_{\odot}}$ 
and 30000 K, respectively. 
Mass of central star and initial mass of disk were set to be 1 $\mathrm{M_{\odot}}$ and 0.01 $\mathrm{M_{\odot}}$, respectively.  
Figure \ref{fig:hydro} shows number density and velocity field at $t = 4.5$ kyr. 
Ultra violet radiation enters computational domain from bottom boundary and propagates along the z-axis. 
We calculated density distribution of dust assuming gas-to-dust mass ratio of 0.01, and 
used it as input data for radiative transfer simulations.   

\begin{figure}[tb]
  \centering
  \includegraphics[width=80mm]{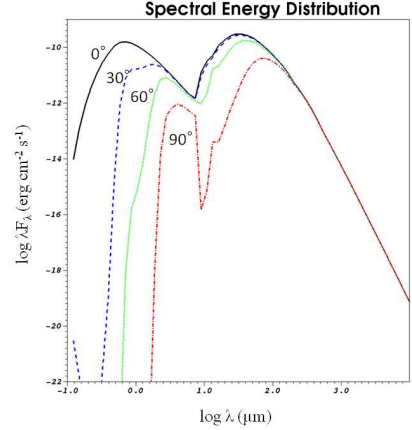}
  \caption{Spectral energy distributions calculated with viewing angle of 0$^\circ$ (solid), 
      30$^\circ$ (dashed), 60$^\circ$ (dotted), and 90$^\circ$ (dash-dotted).}
  \label{fig:SED}
\end{figure}

\begin{figure}[tb]
  \centering
  \includegraphics[width=80mm]{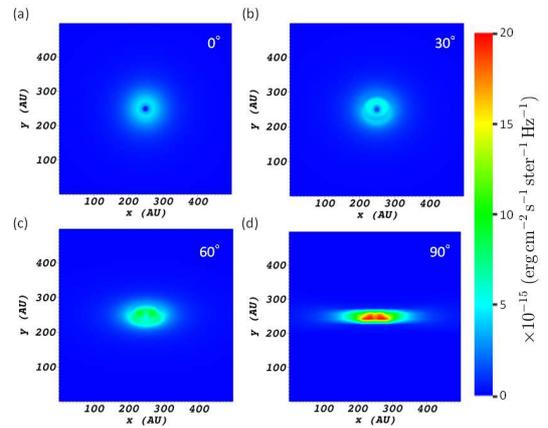}
  \caption{Contour maps of dust continuum intensity at 1000 $\mathrm{\mu m}$ with viewing angle of 0$^\circ$ (a), 
      30$^\circ$ (b), 60$^\circ$ (c), and 90$^\circ$ (d).}
  \label{fig:intensity}
\end{figure}

Figure \ref{fig:SED} shows some selected spectra obtained by our system.
Viewing angle is defined as angle between direction to observer and the positive z-axis. 
Therefore, spectra with viewing angle of 0$^\circ$ are that expected when disk is observed face on. 
As shown in this figure, increasing of viewing angle decreases observed energy flux at short wavelength.   
This is because emission from hot inner region of the disk is obscured by cold outer region of the disk 
when observing edge on disk.
Figure \ref{fig:intensity} shows some selected intensity maps at 1000 $\mathrm{\mu m}$ obtained by our system.
Increasing of viewing angle increases observed specific intensity, because dust continuum at this wavelength 
reflects column density along observed direction.

\section{Summary}\label{sec:summary}
In this paper, we proposed the system for simulating dust continuum observations on distributed computing 
resources. 
Computational tasks and data transfer are automated with RENKEI-WFT in our system.
Our work is important contribution because capability of parallel job execution of our system enables us 
to conduct effective research by radiative transfer simulation. 

Although we focused on simulations of dust continuum observations in this paper, simulations of spectral line 
observations can be implemented in our system with a similar manner. 
Next step would be extension of our system to simulations of spectral line observations.

\section*{Acknowledgment}
This work is partially supported by the Ministry of Education, Culture, Sports, Science and Technology in Japan.

\bibliographystyle{unsrt}

\begin{thebibliography}{1}

\bibitem{1979rpa..book.....R}
G.~B. {Rybicki} and A.~P. {Lightman}.
\newblock {\em {Radiative processes in astrophysics}}.
\newblock Wiley-Interscience, 1979.

\bibitem{1999A&A...344..282L}
L.~B. {Lucy}.
\newblock {Computing radiative equilibria with Monte Carlo techniques}.
\newblock {\em Astronomy and Astrophysics}, 344:282--288, April 1999.

\bibitem{2001ApJ...554..615B}
J.~E. {Bjorkman} and K.~{Wood}.
\newblock {Radiative Equilibrium and Temperature Correction in Monte Carlo
  Radiation Transfer}.
\newblock {\em The Astrophysical Journal}, 554:615--623, June 2001.

\bibitem{kanazawa2010}
H.~Kanazawa, N.~Onishi, Y.~Mizusawa, T.~Tsunekawa, and H.~Usami.
\newblock {Application Hosting Services for Research Community on Multiple Grid
  Environments}.
\newblock {\em Journal of Convergence Information Technology}, 2010.

\bibitem{tatebe2004}
O.~Tatebe, N.~Soda, Y.~Morita, S.~Matsuoka, and S.~Sekiguchi.
\newblock {Gfarm v2: A Grid File System That Supports High-performance
  Distributed and Parallel Data Computing}.
\newblock In {\em Proc. of the 2004 Computing in High Energy and Nuclear
  Physics (CHEP04)}, 2004.

\bibitem{allcock2003}
W.~Allcock.
\newblock {GridFTP: Protocol Extensions to FTP for the Grid}.
\newblock In {\em OGF Document CFD-R.20}, 2003.

\bibitem{Krishnan:2002}
S.~Krishnan, P.~Wagstrom, and G.~von Laszewski.
\newblock {GFSL}: A {W}orkflow {F}ramework for {G}rid {S}ervices, July 2002.
\newblock draft.

\bibitem{anjomshoaa2008}
A.~Anjomshoaa, F.~Brisard, M.~Drescher, D.~Fellows, A.~Ly, S.~McGough,
  D.~Pulsipher, and A.~Savva.
\newblock {Job Submission Description Language (JSDL) Specification, Version
  1.0}.
\newblock In {\em OGF Document CFD-R.136}, 2008.

\bibitem{alfieri2004}
R.~Alfieri, R.~Cecchini, V.~Ciaschini, L.~dell Agnello, \'A. Frohner,
  A.~Gianoli, K.~L\ {o}rentey, and F.~Spataro.
\newblock {VOMS, an Authorization System for Virtual Organizations}.
\newblock In {\em Proc. of the First European Across Grids Conference}, 2004.


\end{thebibliography}

\end{document}